\documentclass[fleqn,12pt,twoside]{article}

\usepackage{espcrc1}

\usepackage{graphicx}
\usepackage{subfig}
\usepackage[countmax]{subfloat}

\usepackage[figuresright]{rotating}

\providecommand{\U}[1]{\protect\rule{.1in}{.1in}}

\newcommand{\AmS}{{\protect\the\textfont2
  A\kern-.1667em\lower.5ex\hbox{M}\kern-.125emS}}

\title{System-Size Dependence in Grand Canonical and Canonical Ensembles}

 \author{Debajit Chakraborty \address[QTP]{Quantum Theory Project, Departments of Physics and of Chemistry,\\ 
University of Florida, Gainesville FL 32611-8435}\thanks{Work supported by U.S.\ Dept.\ of Energy grant DE-SC0002139.} , 
James Dufty \addressmark[Phys]\address{Department of Physics, University of Florida,\\Gainesville, FL 32611, USA}$^*$, and Valentin V. Karasiev \addressmark[QTP]$^*$}

\begin{document}

\maketitle

\tableofcontents
\hfill
\begin{abstract}
The thermodynamics for a system with given temperature, density, and volume is
described by the Canonical ensemble. The thermodynamics for a corresponding
system with the same temperature, volume, and average density is described by
the Grand Canonical ensemble. In general a chosen thermodynamic potential
(e.g., free energy) is different in the two cases. Their relationship is
considered here as a function of the system size. Exact expressions relating
the fundamental potential for each (free energy and pressure, respectively)
are identified for arbitrary system size. A formal asymptotic analysis for
large system size gives the expected equivalence, but without any 
characterization of the intermediate size dependence. More detailed evaluation
is provided for the simple case of a homogeneous, non-interacting Fermi gas.
In this case, the origin of size dependence arises from only two length scales, 
the average inter-particle distance and quantum length scale (thermal deBroglie 
or Fermi length). The free energies per particle calculated from each ensemble 
are compared for particle numbers $2\le N\le 64$ for a range of temperatures above and 
below the Fermi temperature. The relevance of these results for applications 
of density functional theory is discussed briefly.
\end{abstract}\\
   

\section{Introduction and Motivation}

\label{sec1}Equilibrium statistical mechanics provides the fundamental basis
for the thermodynamics of a given system in terms of its Hamiltonian and the
characteristics of its environment (e.g., open or closed) \cite{Hill87}. The
Canonical ensemble applies when the system is in contact with a thermal
reservoir, exchanging energy at constant volume and particle number. It is
parameterized by the temperature ($T\equiv1/k_{B}\beta$),
number density ($n\equiv N/V$), and volume ($V$). The
fundamental thermodynamic potential associated with this ensemble is the
Helmholtz free energy per particle $f_{C}(\beta,n,$ $V)$. The Grand Canonical
ensemble applies under the same thermodynamic conditions but with the additional exchange of
particle number with its environment. It is parameterized by $\beta, \mu,$ and $V$, where $\mu$ is 
the chemical potential. Its thermodynamic potential is the pressure
$p_{G}\left(  \beta,\mu,V\right)  $. However, the free energy per particle in
the Grand Canonical ensemble $f_{G}(\beta,n_{G},V)$ can be determined from
$p_{G}\left(  \beta,\mu,V\right)  $ by a change of variables $\mu\rightarrow
n_{G}\equiv\partial p_{G}/\partial\mu$ via a Legendre transform (see below).
Here $n_{G}$ is the average density in the Grand Canonical ensemble.
Similarly, the pressure can be defined for the Canonical ensemble by the
change of variables 
$n\rightarrow\mu_C\equiv-\partial f_{C}/\partial n$ and
a corresponding Legendre transform.

For large systems it is expected on physical grounds that the system becomes
extensive, in which case the free energy per particle and pressure become
independent of the volume%
\begin{equation}
f_{C}(\beta,n,V)\rightarrow f_{C}\left(  \beta,n\right)  ,\hspace{0.25in}%
p_{G}\left(  \beta,\mu,V\right)  \rightarrow p_{G}\left(  \beta,\mu\right)  .
\label{1.1}%
\end{equation}
Furthermore, if the two ensembles have the same $\beta,V,$ $\ $and $\mu$ is
chosen such that $n=n_{G}(\beta,\mu)$ then the thermodynamics from the two
ensembles should be equivalent in this limit, e.g.
\begin{equation}
f_{C}\left(  \beta,n\right)  =f_{G}\left(  \beta,n_{G}\right)  . \label{1.2}%
\end{equation}

It is this equivalence that allows one to choose an equilibrium ensemble for
convenience of computation or simulation, rather than to fit the actual
experimental conditions of interest. For example, most formulations of density
functional theory are based in the Grand Canonical ensemble while actual
implementations in simulation are for conditions of the Canonical
ensemble, specifically for fixed density and volume. This raises the challenge
of quantifying the conditions for the validity of (\ref{1.1}) and (\ref{1.2}), 
and finding relationships between properties in different ensembles. The objective 
here is to formulate this problem more precisely and to provide some answers
for the simplest case of a non-interacting Fermi gas.

The large system limit is defined by $V\rightarrow\infty$ at constant $n$ or
$n_{G}$ for the Canonical and Grand Canonical ensembles, respectively.
Equivalently, this can be stated as $N\rightarrow\infty$ at constant $n$, or
$N_{G}\rightarrow\infty$ at constant $n_{G}$. In detail, the shape of the
system must be constrained as well, e.g. all dimensions should be of
comparable size $L$ such that $L/r_{0}$ is large, where $r_{0}$ is the average
inter-particle spacing defined by $4\pi nr_{0}^{3}/3=1$. The desired limit
requires that $L$ be large compared to all other characteristic length 
scales as well. One of these is the force range of interaction, $a$. For 
Coulomb systems this is replaced by the screening
length. Another length scale is the thermal de Broglie wavelength $\lambda$
which becomes large at low temperatures, or the corresponding Fermi length
$\widetilde{\lambda}$ at temperatures near zero. There can also be a scale set
by the spatial variations of an external potential. Finally, the correlation
length is typically of the order of the force range, but becomes large near a
critical point so that system-size dependence can be important even for
macroscopic systems. In cases for which  $L$ \ is not the dominant length scale 
the system is ``small" and, while the thermodynamic formalism is universal, the
details must account for the specific environment of the system being
described \cite{Hill}. Here only the Canonical and Grand Canonical conditions
are considered, although many other ensembles for other environments are of
experimental interest \cite{Hill,Hill87}. There is a large literature on the
asymptotic evaluation of the difference between properties calculated in
different ensembles, e.g., fluctuations in extensive variables
\cite{Munster69}. Much less is known away from such asymptotic conditions.
However, low temperature thermodynamic properties of interacting fermions
in 1-D system have also been discussed in the literature (see Ref.\cite{Kurt}).

The next section defines the ensembles and their associated thermodynamic
potentials. In particular, for comparisons exact relationships between them
are identified for arbitrary system size. Generally, the thermodynamic
properties for the two cases are not equal. However, in section \ref{sec3} an
asymptotic analysis for one of these relationships shows their equivalence for
large $V$ (or large $N$) at constant number density. The analysis is formal
and does not expose the full dependence on $V$ nor the cross over to the
extensive limit in (\ref{1.1}). A more detailed quantitative evaluation is
provided in section \ref{sec4} for the special case of a homogeneous non-interacting
gas. In that case, the only relevant length scales are $r_{0}$ and $\lambda$
(or $\widetilde{\lambda}$). Finally, inhomogeneous non-interacting systems
with an external potential are discussed in section \ref{sec5} and related to
the results of section \ref{sec4} using a local density approximation (see
below). The relevance for ensemble dependence and system-size corrections to
the familiar Thomas-Fermi approximation in density functional theory 
\cite{Mermin} is discussed.

\section{Canonical and Grand Canonical Ensembles and Their Thermodynamics}

\label{sec2}The equilibrium Canonical ensemble for a system of $N$ particles
in a volume $V$, coordinates $\mathbf{q}_{i}$, with pairwise interactions and 
an external single particle potential is defined by the probability density operator
\begin{equation}
\rho_{C}=e^{-\beta\left(  H_{N}-Nf_{C}\right)  },\hspace{0.25in}\beta
f_{C}=-\frac{1}{N}\ln Tr_{N}\,e^{-\beta H_{N}}. \label{2.0}%
\end{equation}
Here, $H_{N}$ is the Hamiltonian operator for $N$ particles
\begin{equation}
H_{N}=K_{N}+\Phi_{N}+\sum_{i=1}^{N}v\left(  \mathbf{q}_{i}\right)  ,
\label{2.1}%
\end{equation}
where $K$ and $\Phi$ are the total kinetic and potential energies,
respectively. The specific forms of the pair potential $\phi\left(
\mathbf{q}_{i},\mathbf{q}_{j}\right)  $ and external potential $v\left(
\mathbf{q}_{i}\right)  $ are not required at this point. The equilibrium
thermodynamics for this system is defined from the free energy per particle
$f_{C}(\beta,n,V)$ which is a function of the temperature $T=1/k_{B}\beta$,
the density $n=N/V$, and the volume $V$. The trace in the definition of
$f_{C}$ is taken over the $N$ particle Hilbert space with the appropriate
symmetrization (Bosons or Fermions). For large systems (i.e., $V\rightarrow
\infty$ at fixed finite $n$) it is expected that $f_{C}(\beta,n,V)$ becomes
independent of $V$.

The corresponding Grand Canonical ensemble is \ defined by the operator%
\begin{equation}
\rho_{G}=e^{-\beta\left(  H_{N}-\mu N+p_{G}V\right)  },\hspace{0.25in}\beta
p_{G}V=\ln\sum_{N=0}^{\infty}Tr_{N}\,e^{-\beta\left(  H_{N}-\mu N\right)  }.
\label{2.2}%
\end{equation}
The thermodynamics now is defined from the pressure $p_{G}(\beta,\mu,V)$,
where the density dependence of the Canonical ensemble is replaced by a
dependence on the chemical potential $\mu$. For large systems (i.e.,
$V\rightarrow\infty$ at fixed finite $\mu$) it is expected that the pressure
becomes independent of $V$.

Although the pressure is the fundamentally defined thermodynamic potential in
the Grand Canonical ensemble, the\ corresponding free energy, $f_{G}%
(\beta,n_{G},V),$ is defined in terms of that pressure by a change of
variables from $\mu$ to $n_{G}$ using the Legendre transformation%
\begin{equation}
f_{G}n_{G}=-p_{G}+\mu n_{G}. \label{2.3}%
\end{equation}
Here the average number density, $n_{G}(\beta,\mu,V),$ is%
\begin{equation}
n_{G}(\beta,\mu,V)\equiv\frac{\partial p_{G}(\beta,\mu,V)}{\partial\mu}.
\label{2.4}%
\end{equation}
Similarly, although the free energy is the fundamental potential in the
Canonical ensemble, the pressure $p_{C}\left(  \beta,\mu_{C},V\right)  $ is
defined in terms of that free energy by a change of variables from $n$ to
$\mu_{C}$ using the Legendre transformation%
\begin{equation}
p_{C}=-f_{C}n+\mu_{C}n, \label{2.5}%
\end{equation}
where the chemical potential in the Canonical ensemble is%
\begin{equation}
\mu_{C}(\beta,n,V)\equiv-\frac{\partial f_{C}(\beta,n,V)}{\partial n}.
\label{2.6}%
\end{equation}

From the forgoing definitions it is seen that the thermodynamics defined by the
two ensembles are related exactly by the relation%

\begin{equation}
e^{\beta p_{G}(\beta,\mu,V)V}=\sum_{N=0}^{\infty}e^{\beta\mu N}e^{-\beta
f_{C}(\beta,n,V)N}. \label{2.7}%
\end{equation}
The volume is the same for each term in this summation, so the density $n$
changes accordingly. The inversion of this relationship is obtained in the
Appendix:%
\begin{equation}
e^{-\beta f_{C}(\beta,n,V)N}=\frac{1}{2\pi}\int_{0}^{2\pi}d\theta e^{i\theta
N}e^{\beta p_{G}(\beta,\mu=-i\theta/\beta,V)V} \label{2.8}%
\end{equation}
Note that the Grand Canonical pressure must be analytically extended to
complex values of the chemical potential.

As noted above, the determination of Canonical ensemble properties from given
Grand Canonical ensemble results is relevant for practical applications of
density functional theory. The inversion of (\ref{2.7}) has been discussed
recently \cite{Heras14} where it is proposed to construct $f_{C}(\beta,n,V)$
from a set of linear equations obtained from evaluation of $p_{G}(\beta
,\mu,V)$ at $M$ discrete values of $\mu$. In principle, this requires
$M\rightarrow\infty$ but approximate values for $f_{C}(\beta,n,V)$ are
obtained for finite $M$. More systematic expansions are described in
references \cite{Kosov08,Gonzalez98}. This latter work has been generalized by
Lutsko \cite{Lutsko14}. Equation (\ref{2.8}) appears to be new. A similar
complex expression is given in reference \cite{Munster69}, section 3.2, but
under the assumption that the discrete summation over $N$ in (\ref{2.7}) can
be replaced by an integration. In that case it becomes a Laplace transform for
which the complex Bromwich integral provides its inversion. The explicit
construction of (\ref{2.8}) for the actual discrete case is given in Appendix A.

\section{Thermodynamic Equivalence for Large Systems}

\label{sec3}In this section the limit of large systems is considered. For
it the Canonical and \newline Grand Canonical thermodynamics are expected to
be equivalent. For the Canonical ensemble large systems means the limit
$N\rightarrow\infty$ at constant finite density $n=N/V$ and temperature. For
the Grand Canonical ensemble this limit is $V\rightarrow\infty$ at constant
chemical potential $\mu$ and temperature. To show this equivalence consider
again (\ref{2.8}) written as%

\begin{equation}
\beta f_{C}(\beta,n,V)N=-\ln\frac{1}{2\pi}\int_{0}^{2\pi}d\theta
e^{VA(z=\theta)}, \label{3.1}%
\end{equation}
where now $A(z)$ is a real function of the complex variable $z,$
\begin{equation}
A(z)=izn+\beta p_{G}(\beta,-iz/\beta,V). \label{3.2}%
\end{equation}
It has a stationary saddle point at the value $z\equiv z_{s}$ defined by
$dA/dz=0$. Using (\ref{2.4}) this is
\begin{equation}
\mathrm{Re}\hspace{0.02in} n_{G}(\beta,-iz_{s}/\beta,V)=n,\hspace{0.25in}%
\mathrm{Im}\hspace{0.02in} n_{G}(\beta,-iz_{s}/\beta,V)=0. \label{3.3}
\end{equation}
Since $n_{G}(\beta,z,V)$ is a real function of $z$ the solution is
$z_{s}=i\beta\mu_{s}$ with real $\mu_{s}$ determined from%
\begin{equation}
n_{G}(\beta,\mu_{s},V)=n. \label{3.4}%
\end{equation}

Now let $C$ denote a closed contour in the $z$ plane including the interval
$\left[  0,2\pi\right]  $ along the positive real axis and passing through
$i\beta\mu_{s}$ on the complex axis. Assume that $A(z)$ is analytic on and
within $C$, so that the integral of $\exp(VA(z))$ over the entire contour must
vanish. Consequently, the integral of (\ref{3.1}) can be replaced by an
integration over that part of $C$ complementary to the interval $\left[
0,2\pi\right]  $. Denoting that part by $C^{\prime}$%
\begin{equation}
\beta f_{C}(\beta,n,V)N=-\ln\frac{\beta}{2\pi}\int_{C^{\prime}}dze^{VA(z)},
\label{3.5}%
\end{equation}
where by definition $C^{\prime}$ passes through the stationary point $\mu_{s}$
tangent to the complex axis. Since $A(z)$ is multiplied by $V$, the
contribution near $i\beta\mu_{s}$ \ gives the dominant contribution for large
system size. The usual saddle-point analysis then leads to the asymptotic
result%
\begin{eqnarray}
\beta f_{C}(\beta,n,V) N & \rightarrow &-A(\mu_{s})V-\ln\frac{\beta}{2\pi}%
\int_{-\infty}^{\infty} dx e^{-\frac{1}{2}V\left\vert A^{\prime\prime}
\right\vert \left(x-\mu_{s}\right)^{2}}\nonumber\\
& = & -\mu_{s}\beta N+\beta p_{G}(\beta,\mu_{s},V) V + O(\ln N)\label{3.6}%
\end{eqnarray}
The first two terms are proportional to the free energy of the Grand ensemble
evaluated at the value of the chemical potential that ensures $N_{G}(\beta
,\mu_{s},V)=N$. The free energies are therefore the same up to small
corrections of the order $\left(  \ln N\right)  /N$%
\begin{equation}
f_{C}(\beta,n,V)=f_{G}(\beta,n_{G},V)+O (\frac{1}{N}\ln N) \label{3.7}%
\end{equation}
This is the expected equivalence for large systems. Note, however, that the
analysis does not show that the free energy per particle is independent of
$V$. That question is explored in more detail in the next two sections.

\section{Non-interacting, Homogeneous Systems at Finite System Size}

\label{sec4}In this section the thermodynamics for the Grand Canonical and
Canonical ensembles are calculated exactly at arbitrary system size for the
simplest case of non-interacting particles without external potential. The
Hamiltonian for $N$ particles is%
\begin{equation}
H_{N}^0=\sum_{i=1}^{N}\frac{\widehat{p}_{i}^{2}}{2m}. \label{4.1}%
\end{equation}
For the Canonical ensemble the particle number and volume are fixed so the
boundary conditions chosen here are a cubic box of sides $L$ with hard walls.
Then the momentum components have eigenvalues%
\begin{equation}
p_{\alpha}=\frac{\pi\hbar}{L}k_{\alpha},\hspace{0.25in}\alpha=x,y,z \label{4.2}%
\end{equation}
where $k_{\alpha}$ is a positive integer. The Grand Canonical ensemble represents
an open system without fixed particle number. However, its derivation
represents this as the sum of probabilities for closed systems at the same
volume but different particle number. Hence the same boundary conditions can
be used for calculation of components of each $N$ within the ensemble.

\subsection{Grand Canonical Ensemble}

The pressure in the Grand Canonical Ensemble is given by (\ref{2.2}). 
Since $H_{N}$ is the sum of single-particle operators the summation and trace can be performed directly in
occupation number representation with the result for spin $1/2$ Fermions \cite{Mazenko}
\begin{equation}
\beta p_{G}=\frac{2}{V}\sum_{\mathbf{k}}\ln\left(  1+e^{\beta\mu}e^{-\left(
\frac{k}{\ell}\right)  ^{2}}\right)  . \label{4.3}%
\end{equation}
The three-fold summation is over $\mathbf{k}=k_{x},k_{y},k_{z}$. Use has been
made of%
\begin{equation}
\beta\frac{p^{2}}{2m}=\frac{\beta}{2m}\left(  \frac{\pi\hbar}{L}\right)
^{2}k^{2}=\left(  \frac{k}{\ell}\right)  ^{2},\hspace{0.25in}\ell^{2}=\frac
{4}{\pi}\left(  \frac{L}{\lambda}\right)  ^{2}, \label{4.4}%
\end{equation}
where $\lambda=\left(  2\pi\beta\hbar^{2}/m\right)  ^{1/2}$ is the thermal de
Broglie wavelength. Similarly, the average number density $n_{G}$ is%
\begin{equation}
n_{G}=\frac{2}{V}\sum_{\mathbf{k}}\left(  e^{-\beta\mu}e^{\left(
k/\ell\right)  ^{2}}+1\right)  ^{-1}. \label{4.5}%
\end{equation}

It is tempting at this point to represent the summations over\ $\mathbf{k}$ as
integrals, i.e.
\begin{equation}
\sum_{k_{x}}F\left(k/\ell\right)=\ell\sum_{x}\Delta x F\left(x\right)%
\stackrel{?}{\rightarrow}\ell\int dx F\left(x\right)\label{4.6}%
\end{equation}
Indeed this replacement leads to the familiar textbook results in terms of
Fermi integrals. However, $\Delta x=\Delta k_{x}/\ell=1/\ell$ is small only
for $L/\lambda>>1$. This is not the case for low temperatures or small system
sizes. Hence for the purposes here the discrete summation must be evaluated directly.

At this point all properties will be given a corresponding dimensionless form.
The dimensionless temperature $t$ is%
\begin{equation}
t=\frac{1}{\beta\epsilon_{F}},\hspace{0.2in}\epsilon_{F}=\frac{1}{2m}\hbar
^{2}\left(  3\pi^{2}n_{G}\right)  ^{2/3}. \label{4.7}%
\end{equation}
where $\epsilon_{F}$ is the Fermi energy. It follows that%
\begin{equation}
n_{G}\lambda^{3}=\frac{8}{3\sqrt{\pi}}t^{-3/2}, \label{4.8}%
\end{equation}
so that (\ref{4.5}) becomes%
\begin{equation}
t^{-3/2}=\frac{6}{\pi\ell^{3}}\sum_{\mathbf{k}}\left(  e^{-\beta\mu}e^{\left(
k/\ell\right)  ^{2}}+1\right)  ^{-1}, \label{4.9}%
\end{equation}
and
\begin{equation}
\ell=\left(  \frac{3}{\pi}N_{G}t^{3/2}\right)  ^{1/3}. \label{4.10}%
\end{equation}
An appropriate dimensionless pressure is%
\begin{equation}
p_{G}^{\ast}\left(  t,N_{G}\right)  =\frac{\beta p_{G}}{n_{G}}=\frac{2}{N_{G}%
}\sum_{\mathbf{k}}\ln\left(  1+e^{\beta\mu}e^{-\left(  \frac{k}{\ell}\right)
^{2}}\right)  . \label{4.11}%
\end{equation}
Here it is understood that $\beta\mu=\beta\mu\left(  t,N_{G}\right)  $ as
determined from (\ref{4.9}). Finally, the dimensionless free energy per particle
is obtained from the Legendre transformation as described in (6).
\begin{equation}
f_{G}^{\ast}\left(  t,N_{G}\right)  \equiv-$ $p_{G}^{\ast}\left(
t,N_{G}\right)  +\beta\mu\left(  t,N_{G}\right). \label{4.14}
\end{equation} 

The dimensionless system-size parameter is now $N_{G}$. The analysis proceeds
as follows: 1) choose a value for $N_{G}$ and calculate $\beta\mu\left(
t,N_{G}\right)  $ as a function of $t$ from (\ref{4.9}). Repeat for different
values of $N_{G}$. The results are shown in Figure 1(a). Also shown is the
limiting value for $N_{G}\rightarrow\infty$ obtained from the continuum limit
(i.e., (\ref{4.6}));  2) Calculate $p_{G}^{\ast}\left(  t,N_{G}\right)  $ from
(\ref{4.11}) as a function of $t$ for the same set of values for $N_{G}$. The
results are shown in Figure 1(b);  3) Calculate the dimensionless free energy
$f_{G}^{\ast}\left(  t,N_{G}\right) $ as a
function of $t$ for the same set of $N_{G}$ from (\ref{4.14}). The results are shown in Figure 2.

These figures show that the system-size dependence is small for $N_{G}\geq16$
at $t=10$, but is more significant as the temperature is lowered. This is
expected since that dependence is controlled by $\ell=\left(  \frac{3}{\pi
}N_{G}t^{3/2}\right)  ^{1/3}$ and vanishes only for large $\ell$. Below $t=1$,
larger values of $N_{G }$ are required to approach system-size independence.\\%

\begin{figure}[htp]
\centering
\subfloat[]{\includegraphics[width=7.78cm]{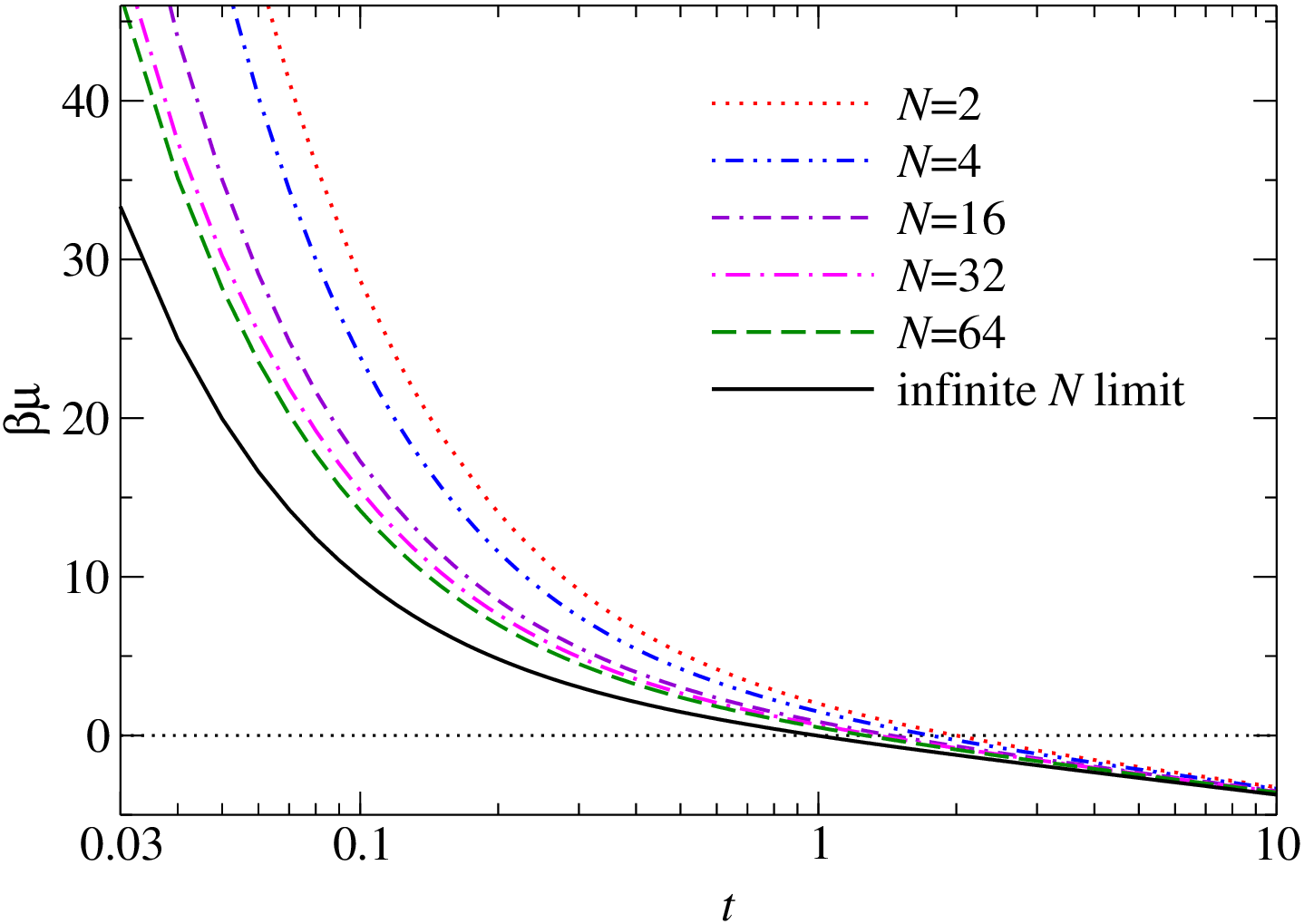}}
\quad
\subfloat[]{\includegraphics[width=7.78cm]{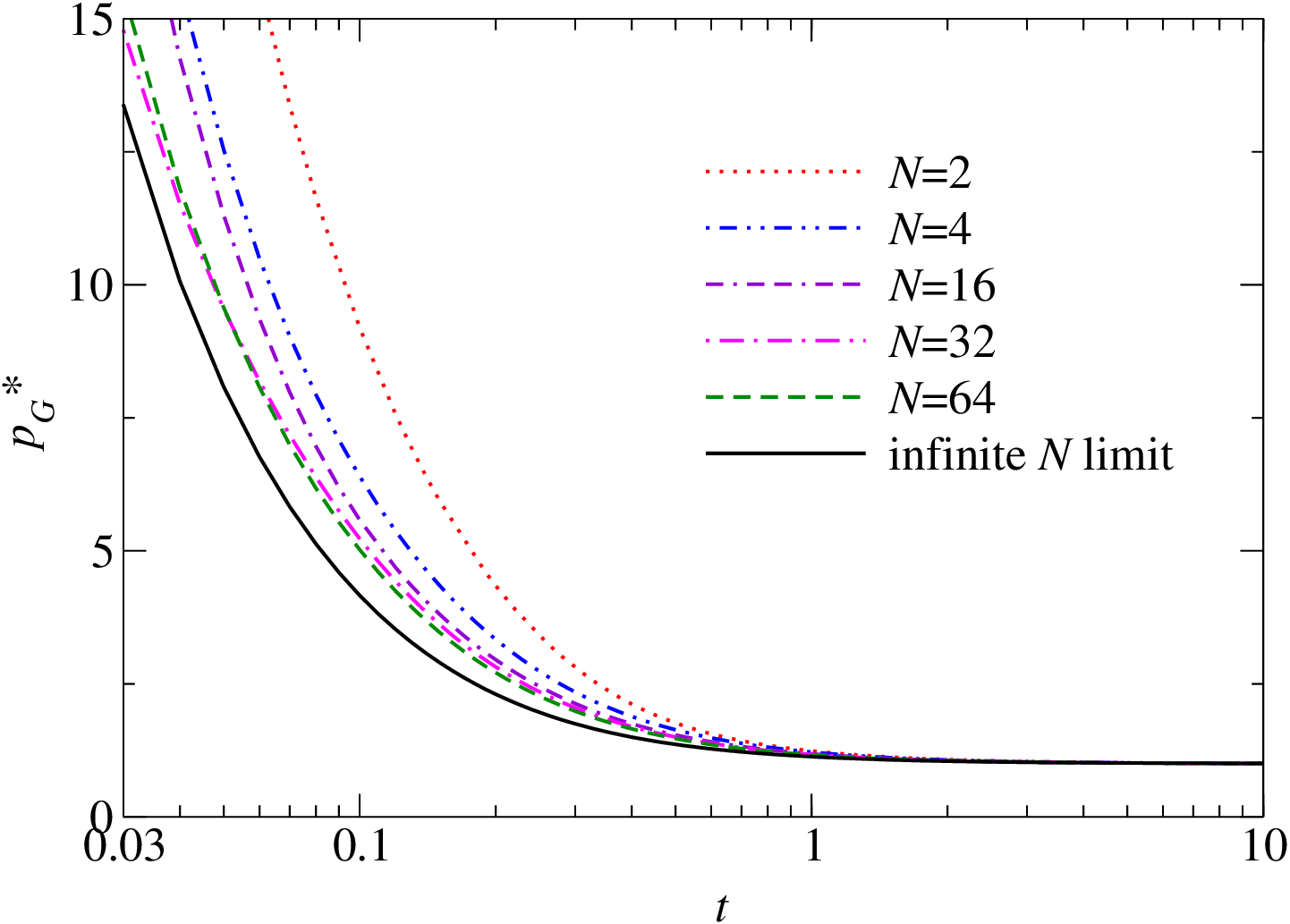}}
\caption{ (a) Plot of $\beta\mu\left(t,N_{G}\right)$ as a function of the
dimensionless temperature $t$ for several values of $N_{G}=N$. Also shown is the large system-size limit.
(b) Plot of the dimensionless pressure $p_{G}^{\ast}\left(t,N_{G}\right)$ as a function of the dimensionless temperature $t$ for
several values of $N_{G}=N$.}
\end{figure}

\begin{figure}[htp]
\centering
\subfloat{\includegraphics[scale=0.45]{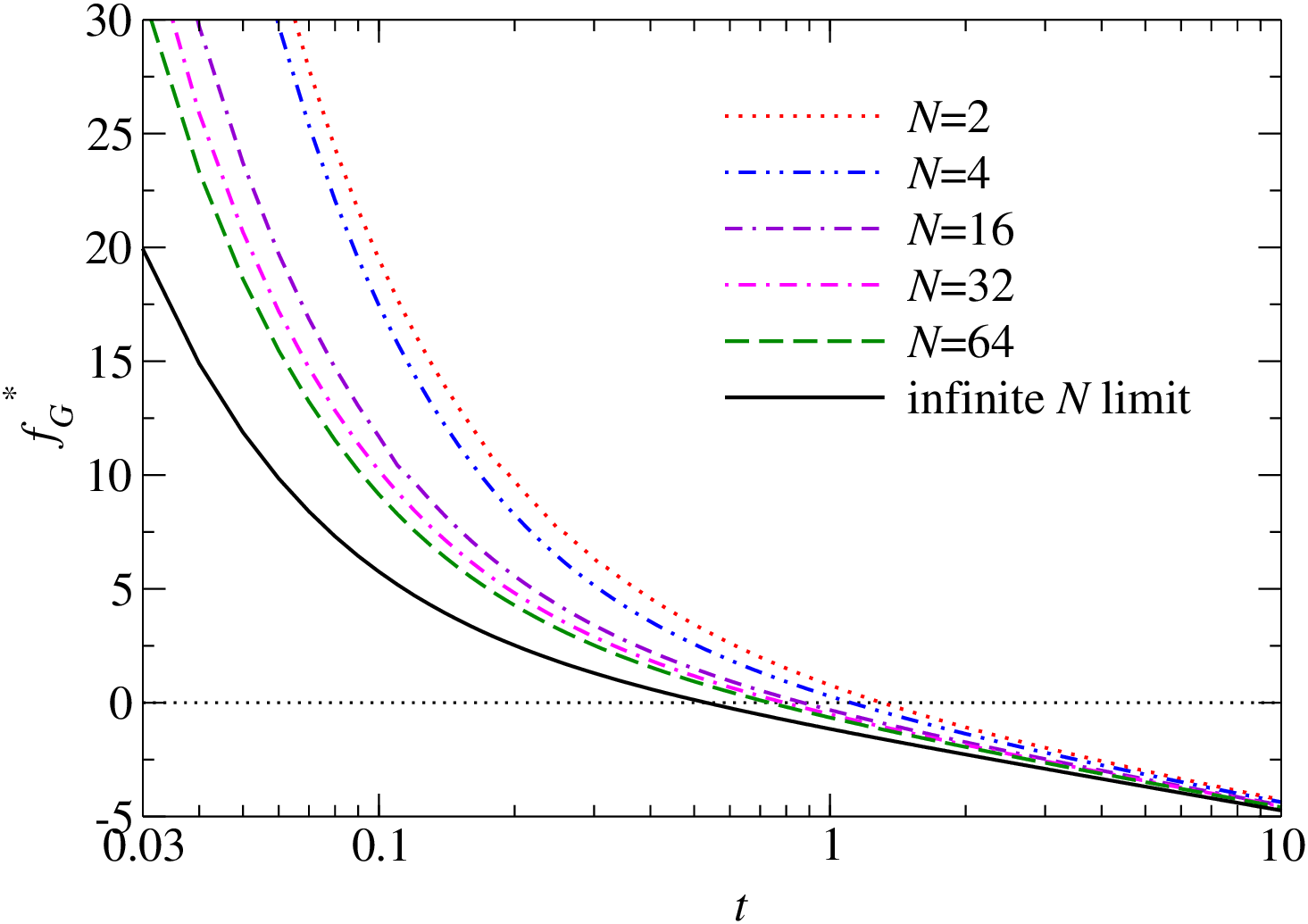}}
\caption{Plot of the dimensionless free energy per particle, $f_{G}^{\ast
}\left(  t,N_{G}\right)=\beta f_G\left( \beta,n_{G},V\right)  $, as a function of the dimensionless temperature $t$ for several values of $N_{G}=N.$}
\end{figure}

\subsection{Canonical Ensemble}

Equation (\ref{2.8}) shows that the Canonical ensemble free energy per
particle can be obtained from the Grand Canonical pressure, extended to
complex values for the chemical potential. It is written as
\begin{equation}
\beta f_{C}(\beta,n,V)=-\frac{1}{N}\ln\frac{1}{2\pi}\int_{0}^{2\pi}d\theta e^{\left(
i\theta+g\left(  \theta,t,N\right)  \right)  N},\label{4.12}%
\end{equation}
with%
\begin{equation}
g\left(  \theta,t,N\right)  \equiv\frac{\beta p_{G}(\beta,\mu=-i\theta
/\beta,V)}{n}=\frac{2}{N}\sum_{\mathbf{k}}\ln\left(  1+e^{-i\theta}e^{-\left(
\frac{k}{\ell}\right)  ^{2}}\right)  .\label{4.13}%
\end{equation}
The definitions of $t$ and $\ell$ are the same as in (\ref{4.7}) and
(\ref{4.10}) except with $n_{G}$ and $N_{G}$ replaced by $n$ and $N$. The
calculation of $g\left(  \theta,t,N\right)  $ is similar to that of
$p_{G}^{\ast}\left(  t,N_{G}\right)  $ in (\ref{4.11}), except that it has
both real and imaginary parts. Their numerical calculation is straightforward
but the final $\theta$ integration of (\ref{4.12}) is now problematical. Due
to the complex integral it has an oscillatory integrand whose variation
increases as $N$, and whose modulation varies between values of the order
$\exp(\pm N)$. Figure $3$ illustrates the problem for $N=64$, $t=0.63$.
This is a precursor for the cross-over to the asymptotic analysis of section
\ref{sec3}. The first difficulty of a rapidly oscillating integrand can be
overcome by increasing the density of mesh points. That works in principle for
both very large $N$ and $t$. The second problem of huge cancellations when the
integral is evaluated numerically as a sum over mesh points does not have a
simple solution because of the finite precision of floating point numbers. As can
be seen from the figure $3$ the highest magnitude of the integrand is $\sim10^{9}$
while the final value of the real part of the integral is $\sim10^{-10}$. Quadruple precision
is required for adequate accuracy within the restricted domain $0.1<t<10$ and
$N\leq64$.%
\begin{figure}[htp]
\centering
\subfloat{\includegraphics[scale=0.4]{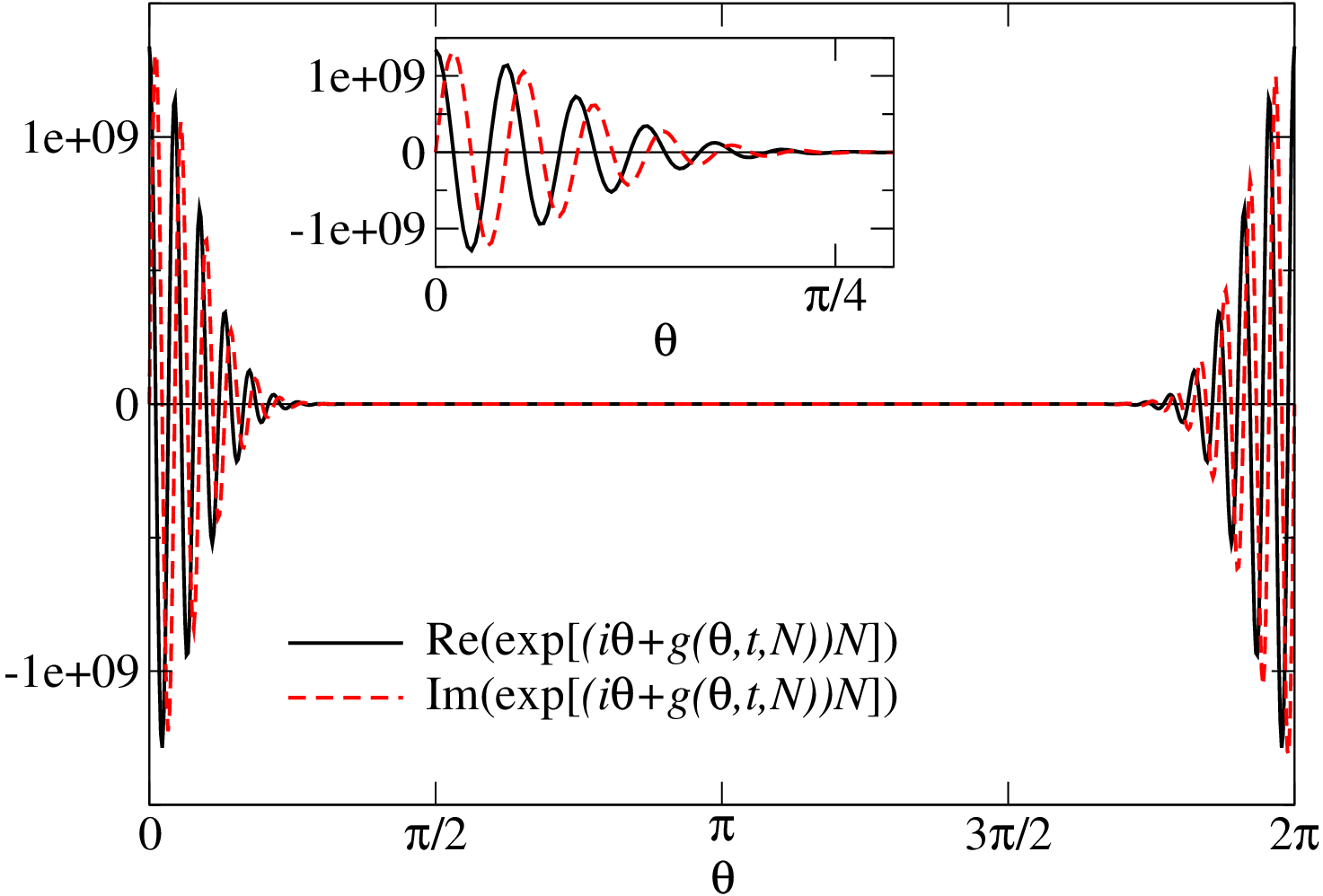}}
\caption{Illustration of the strong variation with $\theta$ for the integrand of (\ref{4.12}), for $N=64$, $t=0.63$}
\end{figure}

\begin{figure}[htp]
\centering
\subfloat[]{\includegraphics[width=7.5cm]{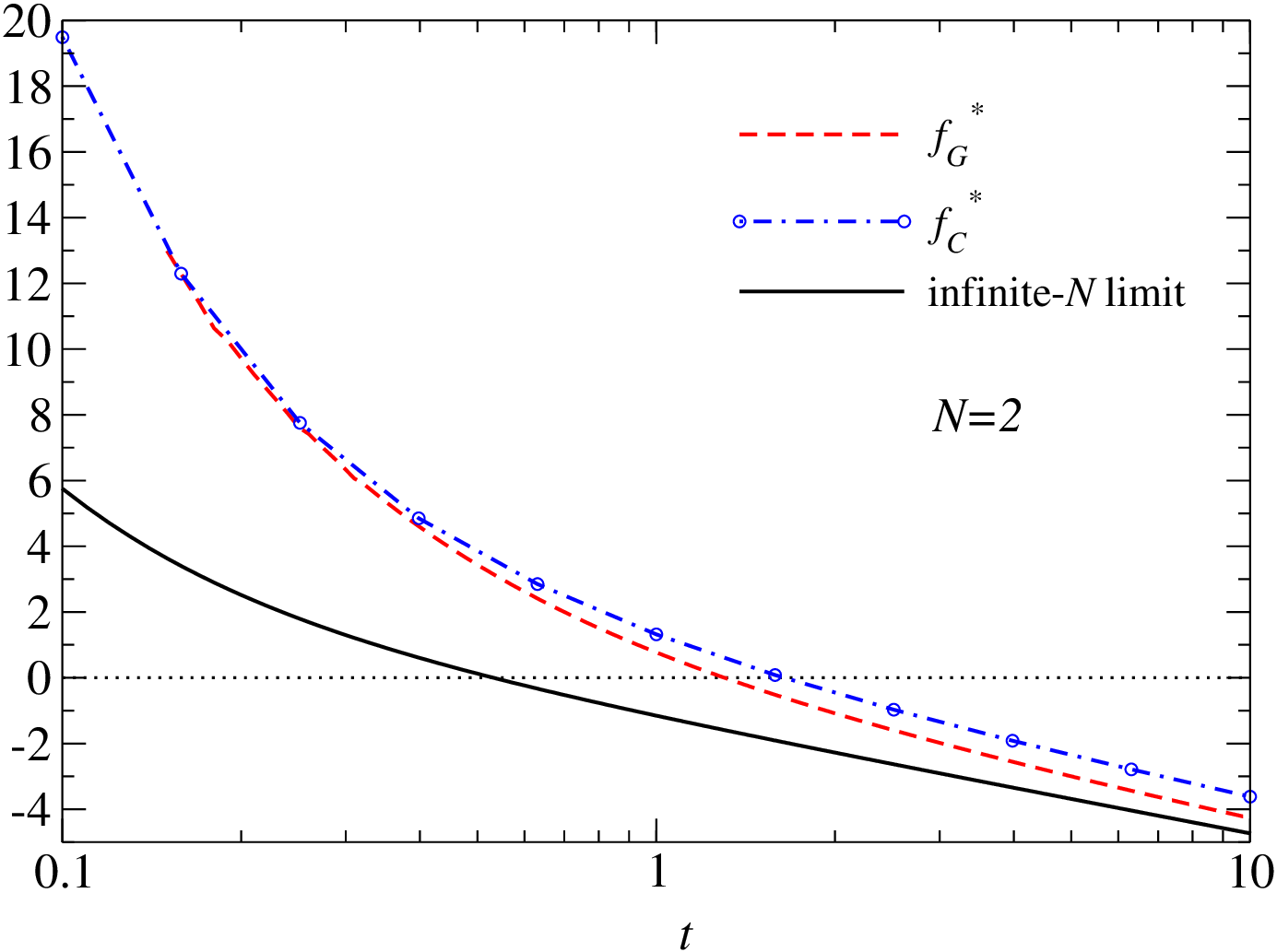}}
\qquad
\subfloat[]{\includegraphics[width=7.5cm]{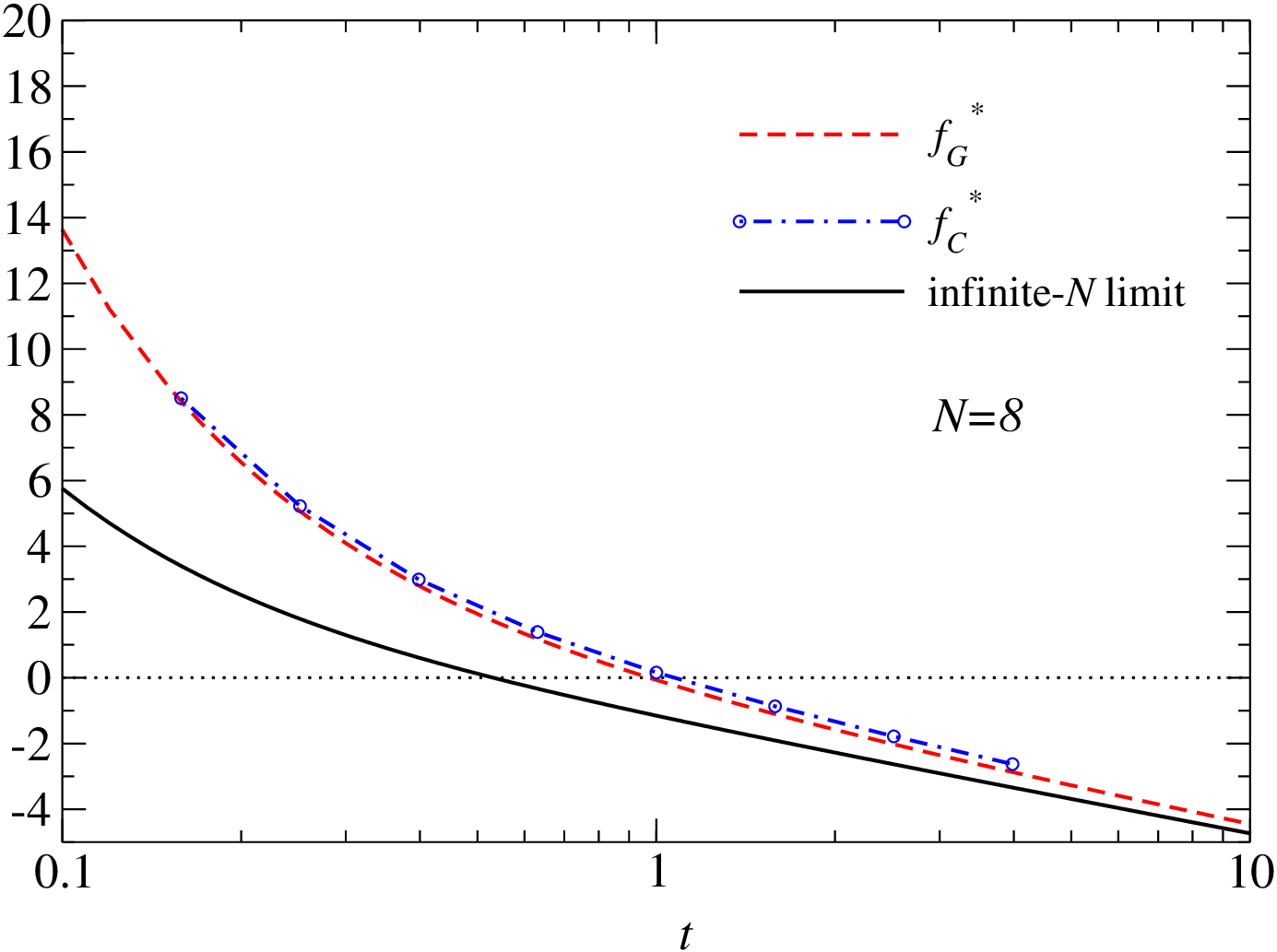}}
\vskip 40pt
\subfloat[]{\includegraphics[width=7.5cm]{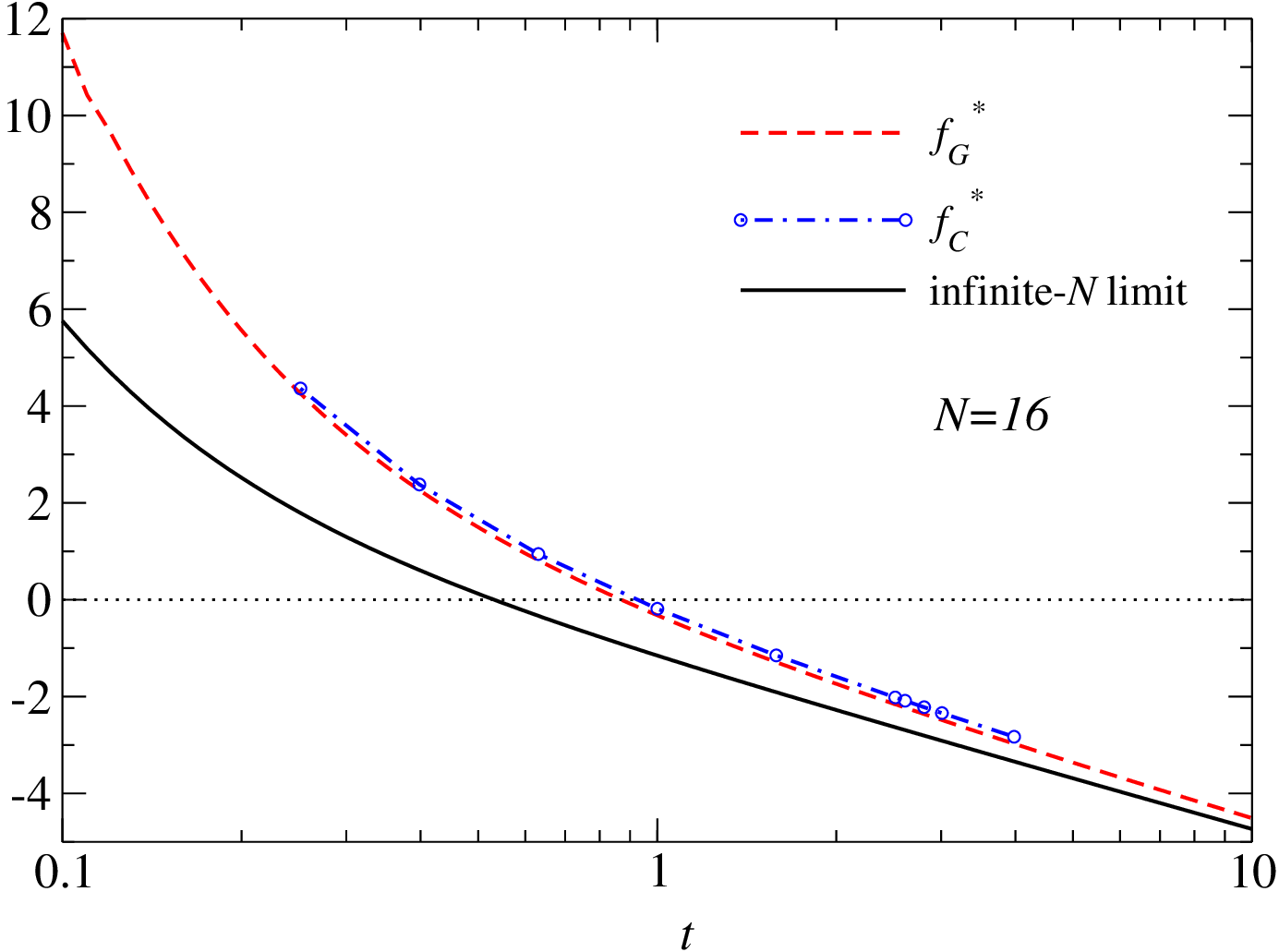}}
\qquad
\subfloat[]{\includegraphics[width=7.5cm]{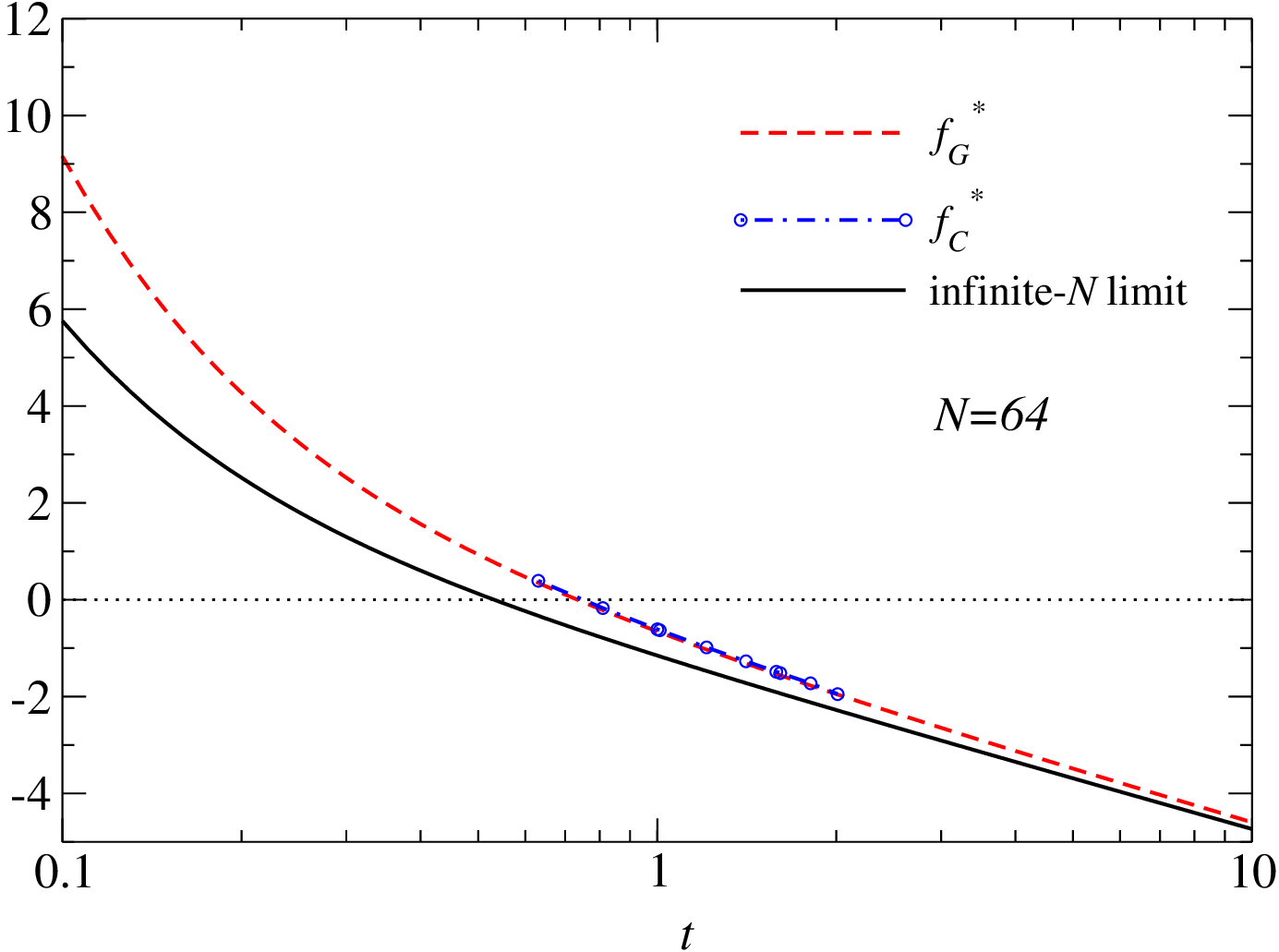}}
\caption{ Panel (a) compares the dimensionless free energy per particle for Canonical and Grand Canonical ensemble $(f_G^*\left(t,N_{G}\right),
f_C^*\left(t,N\right)$ as a function of the dimensionless temperature
$t$ for $N=2$. Also shown is the large system-size limit. Panels
(b)-(d) show the same comparison for $N=8$, $N=16$ and $N=64$ respectively.}
\end{figure}

Figures $4a$ - $4d$ show the dimensionless Canonical ensemble free energy per
particle in comparison with the corresponding Grand Canonical ensemble results
of the last section. Generally, for $N>16$ there is good agreement between the
results of the two ensembles, although significant system-size dependence
relative to the large system-size limit remains. At smaller values of $N$ the
discrepancies between the two ensembles decreases at lower $t$. On the contrary, 
the difference for both ensemble from the large system limit increases with smaller $N$ and $t$. 

For $t=0$ the free energy for both Canonical and
Grand Canonical ensemble are the same for given $N$ as it is the sum of discrete
energies upto the Fermi energy for $N$ particles. This, however, is still different
from the infinite system limit which is an integral over the density of states.

\section{Relationship to Density Functional Theory}

\label{sec5} Density functional theory (DFT) describes the thermodynamics of
an equilibrium, inhomogeneous system whose Hamiltonian has the form
(\ref{2.1}) \cite{Mermin}. The external potential implies that the local
density is non-uniform. DFT has a variational principle that states that the
thermodynamic properties are obtained from a functional of this density at its
extremum. The definition of the functional can be given as follows. First, the
Grand Canonical ensemble pressure and density are computed as functionals of
the external potential as in (\ref{2.2})
\begin{equation}
\beta p_{G}V=\ln\sum_{N=0}^{\infty}Tr_{N}\,e^{-\beta\left(  H_{N}-\mu
N\right)  },\hspace{0.25in}n_{G}(\mathbf{r})\equiv-\frac{\partial p_{G}%
(\beta,\mu,V)}{\partial v\left(  \mathbf{r}\right)  }.\label{5.1}%
\end{equation}
Next, the external potential is eliminated by inverting the second equation to
give $\beta p_{G}V$ as a functional of the density and finally, the density
functional of DFT is then given by  \cite{Mermin}%
\begin{equation}
F_{DFT}\equiv-p_{G}V+\int d\mathbf{r}\left(  \mu-v\left(  \mathbf{r}\right)
\right)  n_{G}\left(  \mathbf{r}\right)  ,\label{5.2}%
\end{equation}
It is understood that the density and external potential in the second term are
now independent functions. They become related by the extremum condition that 
provides the equilibrium density in terms of the external potential. Finally, 
with that relationship established, evaluation of $F_{DFT}$ at its extremum gives 
the Legendre transform  (\ref{2.3}) (extended to the inhomogeneous case) and 
hence the equilibrium Grand Canonical free energy.

It is clear from this brief description of DFT that its theoretical
formulation is tied to the Grand Canonical ensemble. However, in practice
construction of approximate functionals often presumes the large system-size
limit (e.g., Thomas-Fermi and local density approximations). Calculations 
almost always fix the total number of particles, $N$, as in the Canonical 
ensemble. Consequently, system-size corrections and ensemble dependencies are 
overlooked or ignored. The analysis of the previous sections is therefore quite 
relevant for current problems of DFT.

To illustrate this, consider the non-interacting part of the DFT functional
constructed as above%
\begin{equation}
\beta p_{G}^{(0)}V=\int d\mathbf{r}\left\langle \mathbf{r}\left\vert
\ln\left(  1+e^{\beta\mu}e^{-\beta(\frac{\widehat{p}^{2}}{2m}+v(\widehat
{\mathbf{q}}))}\right)  \right\vert \mathbf{r}\right\rangle \label{5.3}%
\end{equation}
\begin{equation}
n_{GC}^{(0)}(\mathbf{r})=\left\langle \mathbf{r}\left\vert \left(
e^{-\beta\mu}e^{\beta(\frac{\widehat{p}^{2}}{2m}+v(\widehat{\mathbf{q}}%
))}+1\right)  ^{-1}\right\vert \mathbf{r}\right\rangle \label{5.4}%
\end{equation}
\begin{eqnarray}
\beta F_{DFT}^{(0)}&= -\int d\mathbf{r}\left\langle \mathbf{r}\left\vert \ln\left(
1+e^{\beta\mu}e^{-\beta(\frac{\widehat{p}^{2}}{2m}+v^{(0)}(\widehat
{\mathbf{q}}\mid n_{GC}))}\right)  \right\vert \mathbf{r}\right\rangle\nonumber\\
&+\int d\mathbf{r}\beta\left(  \mu-v\left(  \mathbf{r}\right)  \right)  n^{(0)}_{GC}(\mathbf{r}%
).\label{5.5}%
\end{eqnarray}
On the right side of (\ref{5.4}) $v^{(0)}(\widehat{\mathbf{q}}\mid n^{(0)}_{GC})$
denotes the inversion of (\ref{5.3}) to obtain
$v\left(  \mathbf{r}\right)  $ as a functional of $n_{GC}^{(0)}(\mathbf{r})$.
Further construction of $\beta F_{DFT}^{(0)}$ is non-trivial for general
external potential and entails diagonalization of the single particle
Hamiltonian $\widehat{p}^{2}/2m+v(\widehat{\mathbf{q}})$ and self-consistent
inversion of the expression for $n_{GC}^{(0)}(\mathbf{r})$ (the Kohn-Sham
approach) \cite{Kohn-Sham,Parr}. A simpler method is the local density approximation
that replaces the operator dependence of the external potential by its value at
the point of interest, $(v(\widehat{\mathbf{q}}))\rightarrow v(\mathbf{r}))$.
Then for instance the density equation can be evaluated in momentum
representation using the same boundary conditions as above %
\begin{equation}
n_{GC}^{(0)}(\mathbf{r})\rightarrow\frac{2}{V}\sum_{\mathbf{k}}\left(
e^{-\beta\left(  \mu-v(r)\right)  }e^{\left(  k/\ell\right)  ^{2}}+1\right)
^{-1}|\psi_{\mathbf{k}}(\mathbf{r})|^2\label{5.6}%
\end{equation}
In the large system-size limit the summation can be represented as an
integration and becomes the familiar finite temperature Thomas-Fermi
approximation%
\begin{equation}
n_{TF}^{(0)}(\mathbf{r})\rightarrow h^{-3}\int d\mathbf{p}\left(
e^{-\beta\left(  \mu-v(r)\right)  }e^{\beta\frac{p^{2}}{2m}}+1\right)
^{-1}\,.\label{5.7}%
\end{equation}
Equations (\ref{5.6}) and (\ref{5.7}) are the same results as for the
homogenous system analysis of the last section, with only the replacement
$\mu\rightarrow\mu-v(r)$. Hence the system-size corrections found there for
small $N,t$ apply here as well, and those corrections for the free energy per
particle identified in Figure 2 and 4(a)-4(d) are required for the DFT functional as well.
Notwithstanding those corrections, it is expected that differences between the
results for the two ensembles are small for $N=N_{G}>16$. %
Further discussion of system-size and ensemble dependence of the DFT 
functional will be given elsewhere.
\label{sec6} 

\section{Acknowledgements}

The authors thank J. Lutsko for sharing unpublished research and for helpful comments.

\appendix

\section{Appendices}
\subsection{Determination of $f_{C}(\beta,n,V)$ from $p_{G}(\beta,\mu,V)$}

\label{apA}The definition of $f_{C}(\beta,n,V)$ in (\ref{2.0}) can be written
in the equivalent form
\begin{equation}
\beta f_{C}=-\frac{1}{N}\ln Tr_{N}\,e^{-\beta H_{N}}=-\frac{1}{N}\ln\sum
_{M=0}^{\infty}Tr_{M}\delta_{N,M}e^{-\beta H_{M}}, \label{A.1}%
\end{equation}
with a representation for the Kronecker delta $\delta_{N,M}$ to get
\begin{eqnarray}
\beta f_{C}(\beta,n,V)  &  = &-\frac{1}{N}\ln\frac{1}{2\pi}\int_{0}^{2\pi
}d\theta e^{i\theta N}\sum_{M=0}^{\infty}Tr_{M}e^{-i\theta M}e^{-\beta H_{M}%
}\nonumber\\
&  = &-\frac{1}{N}\ln\frac{1}{2\pi}\int_{0}^{2\pi}d\theta e^{i\theta N}e^{\beta
p_{G}(\beta,\mu=-i\theta/\beta,V)V}. \label{A.2}%
\end{eqnarray}
The last line follows from the definition of $p_{G}$ in (\ref{2.2}). This
gives the relationship (\ref{2.8}) quoted in the text%
\begin{equation}
e^{-\beta f_{C}(\beta,n,V)N}=\frac{1}{2\pi}\int_{0}^{2\pi}d\theta e^{i\theta
N}e^{\beta p_{G}(\beta,\mu=-i\theta/\beta,V)V}. \label{A.3}%
\end{equation}

The consistency of this result with its inverse (\ref{2.7}) can be
demonstrated by substituting the latter into the right side of (\ref{2.8})%

\begin{eqnarray}
e^{-\beta f_{C}(\beta,n,V)N} &  = &\frac{1}{2\pi}\int_{0}^{2\pi}d\theta
e^{i\theta N}\sum_{M=0}^{\infty}e^{-i\theta M}e^{-\beta f_{C}(\beta
,n=M/V,V)M}\nonumber\\
&  = &\sum_{M=0}^{\infty}\frac{1}{2\pi}\int_{0}^{2\pi}d\theta e^{i\theta\left(
N-M\right)  }e^{-\beta f_{C}(\beta,n=M/V,V)M}\nonumber\\
&  = &e^{-\beta f_{C}(\beta,n,V)N}.\label{A.4}%
\end{eqnarray}

\end{document}